\documentclass[conference]{IEEEtran}
\IEEEoverridecommandlockouts
\usepackage{cite}
\usepackage{amsmath,amssymb,amsfonts}
\usepackage{algorithmic}
\usepackage{graphicx}
\usepackage{textcomp}
\usepackage{xcolor}
\usepackage{booktabs}
\usepackage{booktabs}
\usepackage{multirow}
\def\BibTeX{{\rm B\kern-.05em{\sc i\kern-.025em b}\kern-.08em
		T\kern-.1667em\lower.7ex\hbox{E}\kern-.125emX}}
\begin{document}
	
	\title{Classification of Vocal Bursts for ACII 2022 A-VB-Type Competition using Convolutional Neural Networks and Deep Acoustic Embeddings}

	\author{
		
		\IEEEauthorblockN{Muhammad Shehram Shah Syed}
\IEEEauthorblockA{\textit{Department of Software Engineering} \\
	\textit{Mehran University of Engineering and Technology}\\
	Jamshoro, Pakistan \\
	shehram.shah@faculty.muet.edu.pk}

\and

		\IEEEauthorblockN{Zafi Sherhan Syed}
		\IEEEauthorblockA{\textit{Department of Telecommunication Engineering} \\
	\textit{Mehran University of Engineering and Technology}\\
	Jamshoro, Pakistan \\
	zafisherhan.shah@faculty.muet.edu.pk}
		\and

\IEEEauthorblockN{Abbas Syed}
\IEEEauthorblockA{\textit{Department of Electronic Engineering} \\
	\textit{Mehran University of Engineering and Technology}\\
	Jamshoro, Pakistan \\
zaigham.shah@faculty.muet.edu.pk} }
	
	\maketitle

\begin{abstract}
This report provides a brief description of our proposed solution for the Vocal Burst Type classification task of the ACII 2022 Affective Vocal Bursts (A-VB) Competition. We experimented with two approaches as part of our solution for the task at hand. The first of which is based on convolutional neural networks trained on Mel Spectrograms, and the second is based on average pooling of deep acoustic embeddings from a pre-trained wav2vec2 model. Our best performing model achieves an unweighted average recall (UAR) of $0.5190$ for the test partition, compared to the chance-level UAR of $0.1250$ and a baseline of $0.4172$. Thus, an improvement of around $20\%$ over the challenge baseline. The results reported in this document demonstrate the efficacy of our proposed approaches to solve the AV-B Type Classification task. 
\end{abstract}


\section{Introduction} \label{sec_intro}

Traditionally, automated affect recognition has focused on either visual modality such as facial expressions or audio in terms of speech emotions. Vocal bursts such as gasps, groans, and grunts, which are known to convey affect in human communication, have not been studied in detail but~\cite{Cowen2019a} may be considered one of the seminal works. The Affective Computing and Intelligent Interaction (ACII) Affective Vocal Bursts (A-VB) Competition has introduced four unique challenges that focus on the recognition of affective state from vocal bursts as well as the classification of vocal burst in its various categories. Further details of the A-VB Competition are available in the baseline paper~\cite{Baird2022}.

We participated in the Vocal Burst Type classification task. The organizers provided a large dataset consisting of 59,201 audio recordings that were split into development (training and validation) and test partitions. Each audio file was assigned one of the eight different types of vocal bursts, and challenge participants were asked to propose machine learning models that had been trained on the development partition to predict labels for the test partition. Labels for the test partition were not disclosed to the participants.

In this report, we provide a brief description of our proposed solutions for the Vocal Burst Type classification task. We report that one of our proposed solutions, which is based on Mel Spectrograms and Convolutional Neural Networks, achieved an unweighted average recall (UAR) of $0.4901$ whereas the solution based on deep acoustic embeddings (DAE) achieved a UAR of $0.5190$ for the test partition. When compared to the baseline UAR of $0.4172$, both of our proposed solutions offer a significant improvement to the baseline model, with the DAE approach achieving the best results.

The rest of this report is organized as follows: Section 2 provides a summary of the dataset whereas Section 3 provides a brief description of the process flow diagram for our proposed solution to the task at hand. Experiments and Results for the validation as well as the test partition are discussed in Section 4. We conclude this report in Section 5.

\section{Dataset} \label{sec:dataset} 

The dataset for the A-VB Type classification task consists of 59,201 audio files that total more than 36 hours of audio data. Each audio recording was assigned one of the eight categories (Cry, Gasp, Groan, Grunt, Laugh, Pant, Scream, and Other) that describe the different types of voice bursts. Access to this dataset may be requested through Zenodo~\cite{Cowen2022_AVB_dataset}.

\begin{table}[hp!]
	\centering
	\caption{Summary of class label distribution and training/validation partition audio files}
	\begin{tabular}{@{}lcc@{}}
		\toprule
		\textbf{Class   Label} & \textbf{Training} & \textbf{Validation} \\ \midrule
		Cry & 1,845 & 1,834 \\
		Gasp & 7,104 & 6,844 \\
		Groan & 1,357 & 1,251 \\
		Grunt & 1,348 & 1,322 \\
		Laugh & 4,940 & 4,730 \\
		Pant & 421 & 421 \\
		Scream & 1,573 & 1,590 \\
		Other & 1,366 & 1,393 \\ \bottomrule
	\end{tabular}
\label{tab:label_distribution}
\end{table}

Table~\ref{tab:label_distribution} provides a summary of class label distribution for the training and validation participation and indicates that there is a considerable class imbalance, with some classes (such as Gasp and Laugh) showing up more frequently than others. This tells us that machine learning models for the classification task should contain measures to handle class imbalance.
 
\section{Methodology} \label{sec:methodology}

\begin{figure*}[htp!]
	\centerline{\includegraphics[width=14cm,height=5cm]{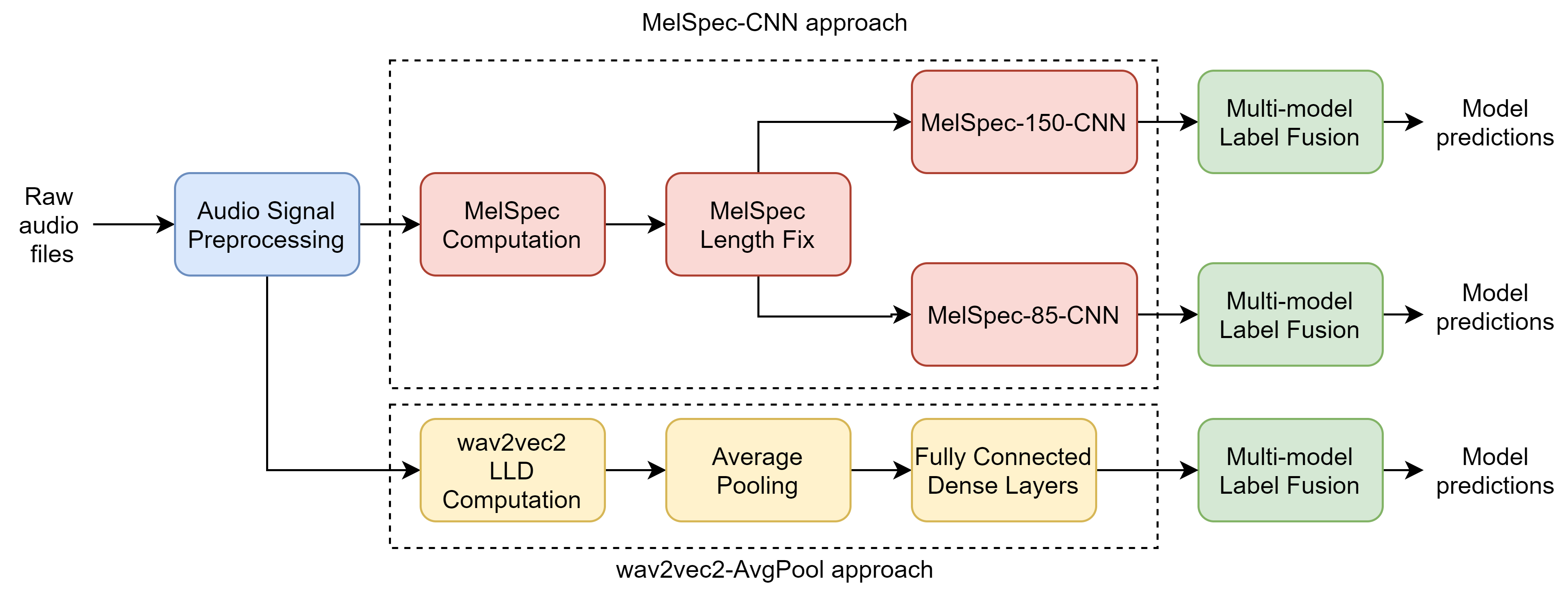}}
	\caption{Process flow diagram of our proposed solutions for ACII-2022 AV-B Vocal Burst Type Classification}
	\label{fig:process_flow}
\end{figure*}

Figure~\ref{fig:process_flow} illustrates the process flow diagram for our proposed solution for the Type classification task. The process begins by applying two types of pre-processing operations for the raw audio files, namely silence removal and amplitude normalization. We explored two different approaches as potential solutions for the classification task. The first approach involved training convolutional neural networks with Mel Spectrogram as input features. The second approach involved computing deep acoustic embeddings from a pre-trained wav2vec2 model and using these as acoustic features after applying average pooling. The final part of our proposed solution used multi-model label fusion to yield improved classification performance. In subsequent sections, we shall briefly describe each step from the process flow diagram.

\begin{figure}[hp!]
	\centerline{\includegraphics[width=7cm,height=7cm]{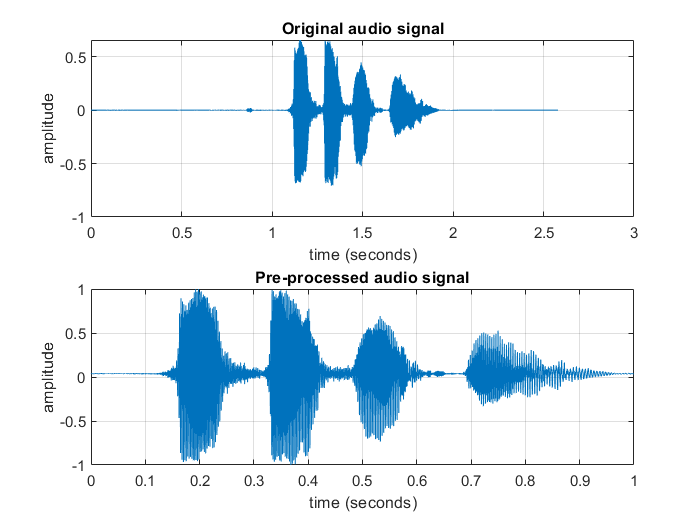}}
	\caption{Time domain graphs of original and preprocessed audio signals (note the difference in x- and y-axes for the two graphs)}
	\label{fig:raw_and_preprocessed_audio}
\end{figure}

We plotted some of the audio signals in the time domain to better comprehend the task at hand and found that these files had a substantial amount of silence at either end of the audio signal. We surmise that such silence does not contribute useful information for the classification task and therefore decided to remove it in an automated manner using the built-in function~\texttt{detectSpeech} in MATLAB. After silence removal, we normalized the amplitude of the audio signal to the range [-1,1]. The effects of silence removal and amplitude normalization are visible in Figure~\ref{fig:raw_and_preprocessed_audio} which plots the original and the preprocessed audio signal in the time domain. It should be noted that even though this method was successful for the vast majority of audio files, some audio files were found to be empty. In such cases, we used predictions from a model (based on the approach described in Section III-B) trained with the extended Geneva Minimalistic Acoustic Parameter Set (eGeMAPS)~\cite{Eyben2016} feature set.

\subsection{Approach 1: Mel Spectrograms and Convolutional Neural Networks}

In this approach, we computed the Mel Spectrogram of audio signals to represent them as images, trained a fully convolutional neural network (CNN) to learn important features from the Mel Spectrograms, and then classified spectrograms into one of the eight classes of vocal bursts.

\begin{figure}[hp!]
	\centerline{\includegraphics[width=7cm,height=7cm]{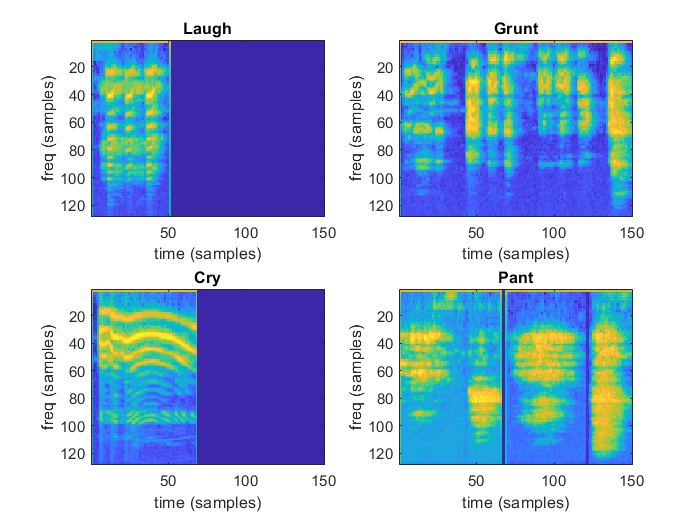}}
	\caption{Examples of Mel Spectrograms for four types of vocal bursts.}
	\label{fig:melspec_examples}
\end{figure}

The Mel Spectrograms were computed using the librosa~toolkit~\cite{McFee2015} for each audio file with the following settings: sampling frequency of 16000 Hz, Mel frequency bins equal to 128, number of DFT bins was set to 512 with a hop length of 256 samples. The magnitude of each Mel Spectrogram was first converted into a logarithmic scale before it was normalized to the range of [0,1]. We used the empirical Cumulative Distribution Function to determine that around $90\%$ of Mel Spectrograms had 150 or fewer temporal bins. Therefore, we decided to fix the length of temporal bins to a constant value of 150 by padding a matrix of zeros to the end of spectrograms if it had a length of less than 150 and to truncate it if the length was greater than 150 temporal bins. We called this setting MelSpec-150. Examples of spectrograms for four types of vocal bursts are illustrated in Figure~\ref{fig:melspec_examples}. In addition to MelSpec-150, we created a MelSpec-85 setting where the length of the spectrogram was fixed to 85 temporal bins. 

We built a relatively shallow CNN to learn meaningful feature representations from these Mel Spectrograms and to subsequently perform classification. The model is organized into three parts, as summarized in Table~\ref{tab:cnn_struct}.

\begin{table}[hp!]
\centering
\caption{Structure of the CNN model used for Mel Spectrogram classification}
\begin{tabular}{@{}ccc@{}}
	\toprule
\multicolumn{3}{c}{Input: (128,150,1) or (128,85,1)} \\ \midrule
\multicolumn{1}{c|}{Conv2D(16,(10,1))} & \multicolumn{1}{c|}{Conv2D(16,(1,10))} & Conv2D(16,(3,3)) \\
\multicolumn{1}{c|}{+ BN + ReLU +} & \multicolumn{1}{c|}{+ BN + ReLU +} & + BN + ReLU + \\
\multicolumn{1}{c|}{SpatialDropout(0.2)} & \multicolumn{1}{c|}{SpatialDropout(0.2)} & SpatialDropout(0.2) \\
\multicolumn{1}{c|}{+ MP (4,2)} & \multicolumn{1}{c|}{+ MP (4,2)} & + MP (4,2) \\ \midrule
\multicolumn{3}{c}{Concatenate} \\ \midrule
\multicolumn{3}{c}{\begin{tabular}[c]{@{}c@{}}BN + Conv2D(32, kernel size= (5,5)) + BN +\\    ReLU + SpatialDropout(0.3) + MP(2,4)\end{tabular}} \\ \midrule
\multicolumn{3}{c}{\begin{tabular}[c]{@{}c@{}}BN + Conv2D(32, kernel size=   (5,5)) + BN +\\  ReLU + SpatialDropout(0.3) + MP(2,4)\end{tabular}} \\ \midrule
\multicolumn{3}{c}{\begin{tabular}[c]{@{}c@{}}BN + Conv2D(64,   kernel size= (3,3)) + BN +\\  ReLU + SpatialDropout(0.4) + MP(2,4)\end{tabular}} \\ \midrule
\multicolumn{3}{c}{BN + Conv2D(16, (3,3)) + GMP} \\ \midrule
\multicolumn{3}{c}{BN + Dense(8) + Softmax} \\ \bottomrule
\end{tabular}
\label{tab:cnn_struct}
\end{table}

A unique characteristic of our proposed model is that three parallel paths are provided to the Mel Spectrograms. These paths form \textit{Part A} of our model. The first path consists of a two-dimensional convolutional layer (Conv2D) with kernels that span across 10 samples in the frequency direction. The Conv2D layer is followed by batch normalization (BN), a rectified linear unit (ReLU), two-dimensional spatial dropout (SpatialDropout2D), and max pooling (MP) of size (4,2). The second path consists of a Conv2D layer with kernels that span across 10 samples in the time direction whereas the third path consists of a Conv2D layer with a generic kernel size of (3,3). The outputs from the three parallel paths are then concatenated before being passed down to the layer below it.

\textit{Part B} of our model consists of four Conv2D layers, the first two of which use a large kernel size of (5,5), and the last two use a kernel size of (3,3). Between these layers, we used batch normalization, ReLU activation, and spatial dropout. Towards the end of Part B, we used global max-pooling (GMP) to create a feature vector that can be used for classification. The final part of the model consists of batch normalization followed by a dense layer with eight units and softmax activation to classify the feature vector into one of the eight classes.

Two models, MelSpec-150-CNN and MelSpec-85-CNN were built using the TensorFlow Keras framework~\footnote{https://www.tensorflow.org/api\_docs/python/tf/keras}. Given the class imbalance in the training partition, data fed into these models were stratified such that an equal number of samples were randomly selected from each label class within a batch. This also meant that training accuracy provided a meaningful measurement of the training progress and computation of UAR for the training partition was not required. We used MixUp-based data augmentation~\cite{Xu2018a} given its noted success for audio classification tasks. In MelSpec-85-CNN but not MelSpec-150-CNN we incorporated further regularization by applying matrix-wrapping across the temporal axis of each spectrogram. Each model was trained using categorical cross-entropy loss, Adam optimization with a learning rate of 0.02, MixUp with $\alpha$ parameter of 0.2, and a pre-MixUp batch size of 400 for a total of 100 epochs. All other parameters were taken as default values in Keras. 

\subsection{Approach 2: Wav2vec2 Embeddings with Fully Connected Dense Layers}
We have previously had success in using pre-trained deep textual embeddings for text classification tasks~\cite{Shehram2020,Syed2021}, therefore, we wanted to explore how well wav2vec2 embeddings~\cite{Baevski2020} would perform for Vocal Burst Type classification. To this end, we computed embeddings from the first four layers of the \texttt{WAV2VEC2\_ASR\_BASE\_960H} pre-trained model available through the torchaudio package~\footnote{https://pytorch.org/audio/stable/index.html}. Like our previous work, we applied average pooling to the embeddings to get a fixed length feature vector of 768 dimensions.

We used a 3-layer fully connected neural network with dense layers as the classifier for the wav2vec2 features. Here, Layer 1 consisted of batch normalization, followed by a Dense layer with 64 units, dropout with a chance of 0.2, and ReLU activation. Layer 2 consisted of a Dense layer with 32 units, dropout, and ReLU activation, and the final layer consisted of a dense layer with 8 units (equal to the number of classes) and softmax activation. Since we did not use data augmentation for this method, we reduced the learning rate to 0.0001 but the model was trained for double the number of epochs i.e., 200 as opposed to 100 epochs for the first approach. All other parameters were set as default values. Also, like Approach 1, we supplied training data in a stratified manner to handle class imbalance. 

\section{Experimental Results}
In this section, we shall briefly describe experimental results for the validation and test partitions.

\subsection{Classification Performance for the Validation Partition}

Table~\ref{tab:val_results} provides a summary of the top performing models from the solutions proposed in Section~\ref{sec:methodology}. Since each training batch was stratified in terms of label distribution, therefore, we use classification accuracy as the performance metric for training partition whereas UAR was used to measure classification performance for the validation and test partitions.

The best models from MelSpec-150-CNN and MelSpec-85-CNN approaches achieved UARs of $0.4834$ and $0.4901$, respectively, for the validation partition, whereas the best result from the wav2vec-AvgPool approach provided a UAR of $0.5098$. All these approaches, therefore, provided better classification performance than chance-level UAR of $0.1250$ and the challenge baseline of $0.4166$ for the validation partition.

\begin{table}[]
	\centering
	\caption{Summary of classification results for the validation partition}
\begin{tabular}{@{}ccc@{}}
	\toprule
	\multicolumn{3}{c}{\textbf{Top-5 models from MelSpec-150-CNN}} \\ \midrule
	\textit{Epoch} & \textit{Train Accuracy} & \textit{Val UAR} \\ \midrule
	81 & 0.5654 & 0.4805 \\
	87 & 0.5677 & 0.4801 \\
	88 & 0.5658 & 0.4798 \\
	91 & 0.5664 & 0.4822 \\
	95 & 0.5674 & 0.4834 \\ \midrule
	\multicolumn{3}{c}{\textbf{Top-5 model from MelSpec-85-CNN}} \\ \midrule
	\textit{Epoch} & \textit{Train Accuracy} & \textit{Val UAR} \\ \midrule
	78 & 0.5658 & 0.4901 \\
	83 & 0.5681 & 0.4841 \\
	89 & 0.5671 & 0.4859 \\
	92 & 0.567 & 0.4861 \\
	93 & 0.5694 & 0.4848 \\ \midrule
	\multicolumn{3}{c}{\textbf{Top-3 models from wav2vec-AvgPool}} \\ \midrule
	\textit{Layer (and Epoch)} & \textit{Train Accuracy} & \textit{Val UAR} \\ \midrule
	Layer 1 (epoch 181) & 0.68 & 0.5031 \\
	Layer 2 (epoch 173) & 0.66 & 0.5092 \\
	Layer 3 (epoch 197) & 0.63 & 0.5098 \\ \midrule
	Chance-level & - & 0.1250 \\
	Challenge baseline & - & 0.4166 \\ \bottomrule
\end{tabular}
\label{tab:val_results}
\end{table}

\subsection{Classification Performance for the Test Partition}
The Vocal Burst Type classification task allowed each participating team to submit five sets of predictions for the Test partition. We will briefly describe our five attempts to predict Test partition labels in this section and present the findings. A summary of results has been provided in Table~\ref{tab:test_results}.

\begin{table}[]
	\centering
	\caption{Summary of classification results from our attempts at predicting test partition labels}
	\begin{tabular}{@{}cc@{}}
		\toprule
		\textbf{Attempt} & \textbf{UAR} \\ \midrule
		1 & 0.4878 \\
		2 & 0.4826 \\
		3 & 0.4968 \\
		4 & 0.4927 \\
		5 & \textbf{0.5190} \\ \midrule
		Chance-level & 0.1250 \\
		Challenge baseline & 0.4172 \\ \bottomrule
	\end{tabular}
\label{tab:test_results}
\end{table}

For attempt 1, we submitted the predictions from the best-performing model from the MelSpec-85-CNN setting. This model achieved a UARs of $0.4901$ and $0.4878$ for the validation and test partitions, respectively. It should be noted that this outcome was better than the challenge baseline.

For our second attempt, we used label fusion of top-5 models in terms of validation partition UAR from MelSpec-85-CNN. We took all five models from the same training loop with the understanding that each model may provide unique insights for predicting labels. This model achieved a test partition UAR of $0.4826$. Interestingly, this attempt yields poorer results than attempt 1, which would suggest that fusing predictions of top-5 models from MelSpec-85-CNN setting is worse than using the single-best model.

For the third attempt, we used label fusion of top-5 models from the MelSpec-150-CNN setting. The final set of predictions for this attempt resulted in a test partition UAR of $0.4968$ which was the best classification performance for the test partition thus far.

Our second attempt had previously shown that the fusion of 5 models yielded poorer results than the single-best model, therefore, for attempt 4, we used a fusion of top-3 models from the MelSpec-150-CNN setting. Since we have had good experience with model fusion in the past, we did not want to employ a single model and instead chose to decrease the number of models that needed to be fused.  It should be noted that the test partition UAR was $0.4968$ when the top-5 models were fused, whereas it only reached $0.4927$ with the fusion of top-3 models. Thus, top-5 model fusion yielded slightly better UAR than top-3 model fusion. These results highlight the data-driven nature of the task at hand where one rule does not apply to all.

For our last attempt, we used label fusion of three models from the wav2vec2-AvgPool approach, choosing one model each from Layer 1, Layer 2, and Layer 3 of the wav2vec model. Interestingly, this method yielded the best performance i.e., UAR of $0.5190$ for the test partition.

\section{Conclusion}

The ACII 2022 Affective Vocal Bursts (A-VB) Competition provided a unique opportunity to perform audio classification for different types of vocal bursts. To this end, we proposed two different approaches to solve the task at hand. The first approach was based on the classification of Mel Spectrograms with a CNN whereas the second approach was based on deep acoustic embeddings (DAE) from a pre-trained wav2vec2 model. Although the MelSpec-CNN approach (UAR = $0.4968$) was able to beat the competition baseline (UAR = $0.4172$) for the Vocal Burst Type classification task, the DAE approach provided us with the best results (UAR = $0.5190$) overall. We believe that our proposed solution offers an interesting perspective on the relatively new application of audio signal processing and machine learning for studying vocal bursts. Codebase for our work is provided in a Github repository accessible via the link \texttt{https://github.com/zss318/ACII\_AVB\_Type}. We plan on writing a more detailed report on our experiments in near future.

\bibliographystyle{IEEEtran}
\bibliography{library}

\end{document}